# PROLIFERATION OF SDDS SUPPORT FOR VARIOUS PLATFORMS AND LANGUAGES

R. Soliday, APS/ANL, Argonne, IL 60439, USA

Abstract

Since Self-Describing Data Sets (SDDS) were first introduced, the source code has been ported to many different operating systems and various languages. SDDS is now available in C, Tcl, Java, Fortran, and Python. All of these versions are supported on Solaris, Linux, and Windows. The C version of SDDS is also supported on VxWorks. With the recent addition of the Java port, SDDS can now be deployed on virtually any operating system. Due to this proliferation, SDDS files serve to link not only a collection of C programs, but programs and scripts in many languages on different operating systems. The platform-independent binary feature of SDDS also facilitates portability among operating systems. This paper presents an overview of various benefits of SDDS platform interoperability.

## 1 VARIOUS PORTS

### 1.1 Versions of C

Originally SDDS programs [1] were written to store and manipulate accelerator data at the Advanced Photon Source (APS). Since that time SDDS usage has spread to many additional facilities with different needs and different hardware. This was accomplished by undertaking the effort to port SDDS programs to additional hardware platforms and to additional programming languages. With the addition of these various ports we have been able to keep a generic source code base that compiles on all the supported operating systems. This has helped to facilitate upgrades because changes in the source code affect all of the SDDS versions.

The SDDS Toolkit, which consists of over 100 applications, was originally written in C on Solaris. A Tcl/Tk extension was also written for Solaris. In order to accommodate other users, the SDDS Toolkit was first ported to the GNU C compiler on Red Hat Linux because it was deemed to be the easiest place to start. This version of the SDDS Toolkit is now available in the form of a binary Red Hat package. It was then ported to Visual C++ and the free Borland C++ Builder on Windows. These versions are now available as self-installing executables. The APS is now also using a VxWorks port of SDDS so that the Experimental Physics and Industrial Control System (EPICS) input/output controllers (IOCs) can read and write SDDS files.

### 1.2 Wrappers and Extensions

Using a FORTRAN wrapper for the SDDS C libraries, it has been possible to incorporate SDDS function calls in FORTRAN programs. A similar technique was used with the Python programming language. Both of these ports require that the SDDS C libraries also be built. This was done because there was a desire to avoid having to maintain multiple source code bases.

### 1.3 Java

Most recently SDDS has been ported to Java. In order to take full advantage of Java's cross-platform capabilities it was decided to avoid using the existing SDDS C libraries. This required writing the Java SDDS port from scratch. This code has evolved over time and is now totally SDDS compatible with the exception of SDDS array types. Benefits of this port include the fact that any Java SDDS application is automatically cross platform and the ability to extend existing Java programs so that they are SDDS compatible.

## 2 APPLICATIONS

### 2.1 SDDS Toolkit and OAG Tcl/Tk Software

Various SDDS applications are distributed through the Operation Analysis Group (OAG) web site[1]. According to our statistics 40% of the downloads are for precompiled Windows software, 20% for precompiled Linux software, and 40% for source code which can be compiled on any of our supported platforms. These applications include the SDDS Toolkit, which is used for postprocessing of the databases [2]. Also included are SDDS EPICS software that can be used to monitor and manipulate process variables (PVs) in the IOCs. There are also various accelerator modeling and simulation programs using SDDS databases.

The original versions of most of the SDDS programs have been refined at the APS on Solaris operating systems. The operators at the APS use C and Tcl/Tk

---

[1] http://www.aps.anl.gov/asd/oag/oaghome.shtml

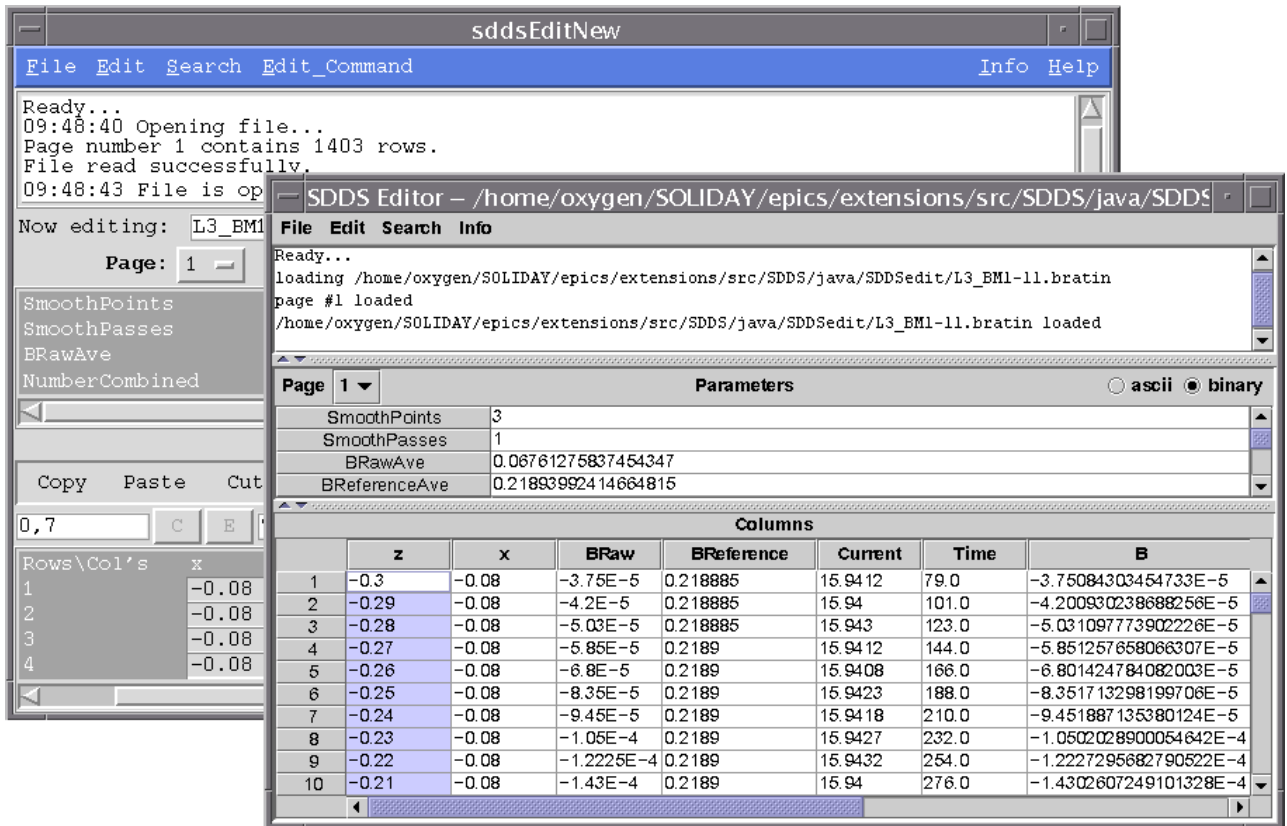

Figure 1: Example of an SDDS file editor written in Java and Tcl/Tk.

SDDS applications extensively. Over time enhancements have been added to existing programs and new programs were written to add further functionality. The usual setup of a program used in the APS control room consists of a Tcl/Tk graphical user interface (GUI) that directly interfaces with SDDS files and PVs or a Tcl/Tk GUI that executes various SDDS C programs. These programs include the Procedure Execution Manager (PEM), which has been used to automate accelerator operations [3,4].

## 2.2 VxWorks Applications

The core SDDS library routines that have also been ported to VxWorks are now being used at the APS to load and store configuration data in some of the IOCs, as well as being used to write video images to disk. Another more limited use of SDDS in the IOCs has been to run some of the SDDS Toolkit applications. If demand for this increases further, SDDS Toolkit applications will be ported to VxWorks.

## 2.3 Accelerator Simulators

As a result of requests from other institutions and the prospect of using faster and less expensive hardware, these programs have been ported to both Windows and Linux. Beyond simply porting these programs, they are also now available in precompiled binaries. This makes deployment of these programs much easier for most users. Currently these ported versions are being used at the APS to run accelerator simulations on very fast computers that are relatively inexpensive. Future plans include investing in more x86 processors to run large-scale simulations.

In addition to porting to other platforms, there has been work done to port to additional languages. A FORTRAN wrapper is used to add SDDS functionality to GENESIS [5], a time-dependent free-electron laser (FEL) simulation code written by Sven Reiche. The code was modified so that the output files can be plotted with existing SDDS plotting programs. Also the output from elegant, an accelerator simulation program, can now be used as input to GENESIS. This work has helped to achieve start-to-end accelerator simulations [6].

## 2.4 Java

Most recently SDDS has been ported to Java. Using this new language, a few new programs have already been created. A cross-platform SDDS file editor (see Figure 1) was written loosely based on a Tcl/Tk SDDS editor. Another program written in Java is a three-dimensional plotting program (see Figure 2). This program can be used to plot SDDS data as a surface plot or a scatter plot. New SDDS applications can be relatively easy to create in Java from scratch or by integrating SDDS functionality with a preexisting program. One obvious advantage to writing programs in Java is that the programs will run on operating systems that are not supported by previous SDDS ports, such as Macintosh.

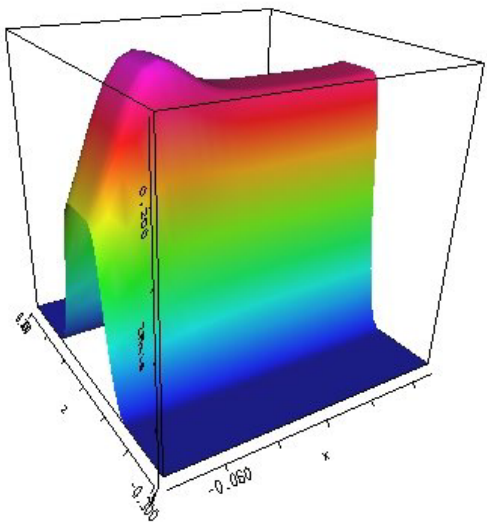

Figure 2: Three-dimensional plot using Java SDDS 3D plotter.

## 3 BENEFITS

### 3.1 Cross Platform

One benefit to having cross-platform database software is that there is no need to change the existing computer infrastructure to use SDDS software. It is not even required that all the computers at one site be of the same type. The SDDS software has binary interoperability, which allows the programs the ability to read SDDS files that were created on different operating systems.

### 3.2 Easy Deployment

Now that there are binary SDDS programs available on Windows and Linux, SDDS is much easier to deploy since the source code does not have to be recompiled. This has brought SDDS software up to the standards of many commercial packages that people have become accustomed to using. The programs are packaged using InstallShield on Windows and RedHat Package Manager on Linux. For the Solaris operating system it is still necessary to compile the source code because this is the standard distribution method for Solaris.

### 3.3 Stable Software

One unexpected advantage of porting SDDS to various operating systems and different C compilers was the ability to detect and remove problems in the source code. Each C compiler has a different set of compiler warnings that we used to refine the SDDS software. With these changes many possible problems have been averted before they became apparent. Using this technique we can check changes to the source code prior to public release.

Automated test scripts written for Solaris, Linux and Windows also test these programs. These scripts run each program through a series of tests using various options. More detailed debugging is accomplished on Solaris by using Purify and Quantify from Rational Software.

## ACKNOWLEDGMENT

Work is supported by U.S. Department of Energy, Office of Basic Energy Sciences, under Contract No. W-31-109-ENG-38.